\documentclass[jmp,aps,showpacs,reprint]{revtex4-1}
\UseRawInputEncoding
\usepackage{graphicx}
\usepackage{dcolumn}
\usepackage{amsmath}
\usepackage{xcolor}
\usepackage{epsfig}
\RequirePackage{xspace}

\usepackage{relsize}

\begin{document}

\title{
\large \bfseries \boldmath Possible solution of the puzzle for the branching ratio and $CP$ violation in $B\to \pi\pi$ decays with a modified perturbative QCD approach}

\author{Sheng L\"{u}} \email{shenglyu@mail.nankai.edu.cn}
\author{Mao-Zhi Yang}\email{yangmz@nankai.edu.cn}
\affiliation{School of Physics, Nankai University, Tianjin 300071, People's Republic of China}

\date{\today}


\begin{abstract}
We study $B\to \pi\pi$ decay with a modified perturbative QCD approach. The branching ratios and $CP$ violation are calculated with the transverse momenta of partons considered. Sudakov factor associated with each meson is included to suppress soft contribution in QCD. With the wave function of $B$ meson obtained in QCD-inspired relativistic potential model being used, the suppression of Sudakov factor to the soft contribution is not effective enough. Soft scale cutoff and soft form factors of $B\pi$ transition and $\pi\pi$ production have to be introduced. The main next-to-leading-order contributions of vertex correction, the quark-loop and magnetic penguin are included. To solve the long-standing puzzle in $B\to\pi\pi$ decay, that is the theoretical prediction of the branching ratio of $B^0\to \pi^0\pi^0$ being seriously smaller than experimental data, color-octet matrix element which is of long-distance dynamics is introduced. With parameters taken with reasonable values, all the branching ratios and $CP$ violation are well consistent with experimental data.
\end{abstract}

\maketitle
\section{Introduction}
High precision data collected by B factories reveal serious discrepancies between experimental measurements and theoretical predictions for branching ratios and $CP$ asymmetry in several $B$ decay modes calculated with perturbative QCD (PQCD) approach \cite{PQCD1,PQCD2,PQCD3} and QCD factorization (QCDF) approach \cite{QCDf1,QCDf2,QCDf3,QCDf4}. For $B\to \pi\pi$ decay, original prediction of the branching ratio of $B\to \pi^0\pi^0$ decay of PQCD is much smaller than experimental data. The branching ratios and $CP$ violation of $B\to \pi\pi$ decays measured in experiment are \cite{PDG2020}
\begin{eqnarray} \label{Br-exp}
&&B(B^+\to \pi^+\pi^0)=(5.5\pm 0.4)\times 10^{-6},\nonumber\\
&&B(B^0\to \pi^+\pi^-)=(5.12\pm 0.19)\times 10^{-6},\\
&&B(B^0\to \pi^0\pi^0)=(1.59\pm 0.26)\times 10^{-6},\nonumber
\end{eqnarray}
and
\begin{eqnarray}
&&A_{CP}(B^+\to \pi^+\pi^0)=0.03\pm 0.04,\nonumber\\
&&A_{CP}(B^0\to \pi^+\pi^-)=0.32\pm 0.04,\\
&&A_{CP}(B^0\to \pi^0\pi^0)=0.33\pm 0.22.\nonumber
\end{eqnarray}
The leading-order prediction for $B(B^0\to \pi^0\pi^0)$ in PQCD is only at the order of $\sim 10^{-7}$ \cite{PQCD3,PQCD4}, which is much smaller than the observed branching ratio in experiment given in Eq. (\ref{Br-exp}), while the predictions for the branching ratios of the other two decay modes are approximately the same order as experimental data. The prediction of QCD factorization approach for the branching ratios of $B\to \pi\pi$ gives similar situation \cite{QCDf1,QCDf2,QCDf3,QCDf4}. This is the so-called $\pi\pi$ puzzle in $B$ decays. The calculation including next-to-leading order contribution of QCD in PQCD still cannot solve $\pi\pi$ puzzle of $B$ decays satisfactorily \cite{PQCD+NL,PQCD+NL2,PQCD+NL3}.

To solve the puzzle, a soft factor associated with pion as an additional nonperturbative input is introduced by analyzing the soft divergence in $k_\text{T}$ factorization in Refs. \cite{Li-Mishima2011,Li-Mishima2014}. The agreement of the theoretical prediction with experimental data is improved, but it is still not in an enough  satisfactory manner.

We reanalyzed the $B\to \pi$ transition form factor in the framework of perturbative QCD approach in Ref. \cite{Lu-Yang2021} using the wave function of $B$ meson obtained by solving bound state equation in QCD inspired relativistic potential model \cite{Yang2012,LY2014,LY2015,SY2017,SY2019}. With the new $B$ meson wave function being used, we find that the suppression of Sudakov factor to the soft contribution in QCD is no longer strongly effective. The soft contribution can be as large as 40\% in the naive calculation. Then a soft momentum cutoff and a soft form factor have to be introduced to make the perturbative calculation reliable. That is the factorization formula for the $B\to\pi$ transition form factor is changed. The dynamics with momentum transfer larger than the cutoff scale can be treated by perturbative calculation, while the contribution lower than the cutoff scale is attributed to the soft form factor.

In this work we study $B\to\pi\pi$ decay in the modified PQCD approach. The branching ratios and $CP$ violation parameters are calculated. For the momentum transfer larger than the soft cutoff scale, the amplitudes are calculated in perturbation theory, and for the interaction lower than the soft cutoff scale, the soft $B\to \pi$ transition and $\pi\pi$ production form factors are introduced, which are treated as nonperturbative inputs. For the perturbative part, the next-to-leading order contribution in QCD are included, the main important contributions of which are from the diagrams of vertex correction, quark-loop diagrams and color magnetic penguin diagrams. By confronting the theoretical results to experimental data, the nonzero color-octet matrix element $\langle\pi\pi |(\bar{q}_1T^aq_2)(\bar{q}_3T^ab)|\bar{B}\rangle$ is introduced, which comes from the color decomposition of the quark-antiquark current operators in the effective Hamiltanion of $B$ decays, and the quark-antiquark current can be either $V\pm A$ or $S\pm P$ currents. We find color-octet matrix element is important to explain the experimental data of $B\to \pi\pi$ decays. By taking reasonable values for the input parameters, we obtain theoretical results for the branching ratios and $CP$ asymmetry parameters in perfect agreement with the data.

The remaining part of the paper is as follows. The leading and next-to-leading order contributions to the perturbative part of the $B\to\pi\pi$ decays are calculated in Sec. II. The contributions of the soft form factors are presented in Sec. III. The contributions of color-octet matrix element are discussed in Sec. IV. The numerical result and discussion are given in Sec. V. Finally, section VI is for a brief summary.

\section{The hard decay amplitude in perturbative QCD}

\subsection{The leading order contribution}

The effective Hamiltonian for the charmless hadronic $B$ decays induced by $b\to d$ transition is \cite{Hamiltanion1996}
\begin{eqnarray}
	\mathcal{H}_{\mathrm{eff}} &=& \frac{G_F}{\sqrt{2}}\bigg[ V_{ub}V_{ud}^*\big(C_1O_1^u+C_2 O_2^u\big)\nonumber \\
	&-&V_{tb}V_{td}^*\bigg(\sum_{i=3}^{10} C_i O_i+C_{8\textsl{g} } O_{8\textsl{g} } \bigg) \bigg],
\end{eqnarray}
where $G_F = 1.16638 \times 10^{-5}~\mathrm{GeV}^{-2}$, is the Fermi constant, $V_{qb}$ and $V_{qd}$ $(q=u,~c,~t)$ the Cabibbo-Kobayashi-Maskawa (CKM) matrix elements, $C_i$'s the Wilson coefficients. The operators in $\mathcal{H}_{\mathrm{eff} }$ are
\begin{eqnarray}
	&&O_1^u = \bar{d}_{\alpha}\gamma^{\mu}L u_{\beta}\cdot \bar{u}_{\beta}\gamma_{\mu}L b_{\alpha},
	 \nonumber  \\
	&&O_2^u = \bar{d}_{\alpha}\gamma^{\mu}L u_{\alpha}\cdot \bar{u}_{\beta}\gamma_{\mu}L b_{\beta},
	\nonumber  \\	
	&&O_3 = \bar{d}_{\alpha}\gamma^{\mu}L b_{\alpha}\cdot \sum_{q'}\bar{q}'_{\beta}\gamma_{\mu}L q'_{\beta},
	\nonumber  \\
	&&O_4 = \bar{d}_{\alpha}\gamma^{\mu}L b_{\beta}\cdot \sum_{q'}\bar{q}'_{\beta}\gamma_{\mu}L q'_{\alpha},
	\nonumber  \\	
	&&O_5 = \bar{d}_{\alpha}\gamma^{\mu}L b_{\alpha}\cdot \sum_{q'}\bar{q}'_{\beta}\gamma_{\mu}R q'_{\beta},
	\nonumber  \\	
	&&O_6 = \bar{d}_{\alpha}\gamma^{\mu}L b_{\beta}\cdot \sum_{q'}\bar{q}'_{\beta}\gamma_{\mu}R q'_{\alpha},
  \\	
	&&O_7 = \frac{3}{2}\bar{d}_{\alpha}\gamma^{\mu}L b_{\alpha}\cdot \sum_{q'} e_{q'} \bar{q}'_{\beta}\gamma_{\mu}R q'_{\beta},
	\nonumber  \\	
	&&O_8 =  \frac{3}{2}\bar{d}_{\alpha}\gamma^{\mu}L b_{\beta}\cdot \sum_{q'} e_{q'} \bar{q}'_{\beta}\gamma_{\mu}R q'_{\alpha},
	\nonumber  \\	
	&&O_9 =  \frac{3}{2}\bar{d}_{\alpha}\gamma^{\mu}L b_{\alpha}\cdot \sum_{q'} e_{q'} \bar{q}'_{\beta}\gamma_{\mu}L q'_{\beta},
	\nonumber
  \end{eqnarray}
  \begin{eqnarray}
  		&&O_{10} =  \frac{3}{2}\bar{d}_{\alpha}\gamma^{\mu}L b_{\beta}\cdot \sum_{q'} e_{q'} \bar{q}'_{\beta}\gamma_{\mu}L q'_{\alpha},  	\nonumber \\
&& O_{8\textsl{g}}=\frac{g_s}{8\pi^2}m_b\bar{d}_\alpha\sigma^{\mu\nu}RT^a_{\alpha\beta}b_\beta G_{\mu\nu}^a, \nonumber	\end{eqnarray}
where $\alpha$ and $\beta$ are the color indices; $L=(1-\gamma_5)$ and $R=(1+\gamma_5)$, are the left- and right-handed projection operators. The sum over $q'$ includes all the quark flavors that are active at $\mu=\mathcal{O}(m_b)$ scale, i.e. $q'\in \{u,d,s,c,b\}$.

If the momentum transfer by gluon is larger than a critical scale $\mu_c$, which separates the hard and soft scales, then the decay process can be treated in perturbation theory of QCD. In general, the value of the critical separation scale is approximately $\mu_c=1.0~\mathrm{GeV}$. For the hard dominant region, i.e., the scale of the gluon momentum is greatly large than $\mu_c$, the hadronic $B$ decay amplitude can be written in a factorized form, where soft interactions associated with each meson can be absorbed into the meson wave functions, and the hard contribution can be calculated as a hard amplitude at quark level. Then the $B$ decay amplitude can be written as a convolution of hard part and meson wave functions
\begin{eqnarray}
&&\mathcal{M}=\int d^3k \int d^3k_1 \int d^3 k_2 \Phi^B(\vec{k},\mu)\nonumber \\
&&\times C(\mu) \times H(k,k_1,k_2,\mu)\times \Phi^\pi (k_1,\mu)\Phi^\pi (k_2,\mu).
\end{eqnarray}
Here $H$ stands for the hard part of decay amplitude, $\Phi^{B,\pi}$ the meson wave functions, and $C(\mu)$ the combination of the Wilson coefficients in the decay.

The wave function of $B$ meson can be defined through the matrix element $\langle 0| \bar{q}(z)_\beta [z,0]b(0)_\alpha |\bar{B}\rangle$ by
\begin{equation}
\langle 0| \bar{q}(z)_\beta [z,0]b(0)_\alpha |\bar{B}\rangle =\int d^3k \Phi^{B}_{\alpha\beta}(\vec{k})e^{-ik\cdot z},
\end{equation}
where the right-hand side of the above equation is written in the rest-frame of $B$ meson, and $[z,0]$ denotes the path-ordered exponential $[z,0]={\cal P}\exp[-ig_sT^a\int_0^1 d\alpha z^\mu A_\mu^a(\alpha z)]$.

If considering the decay in the rest-frame of $B$ meson, only the wave function in the rest-frame of $B$ meson is needed. In this work, we use the wave function of $B$ meson that was derived by solving the bound-state equation in the QCD-inspired relativistic potential model in Refs. \cite{Yang2012,LY2014,LY2015}, where the whole mass spectrum of $b\bar{q}$ system was calculated, and the results are in good agreement with experimental data. The spinor wave function $\Phi^{B}_{\alpha\beta}(\vec{k})$ has been derived in Ref. \cite{SY2017}, which is
\begin{eqnarray} \label{B-wave}
  \Phi_{\alpha\beta}(&\vec{k}&)=\frac{-if_Bm_B}{4}K(\vec{k})
\nonumber\\
 && \cdot\Bigg\{(E_Q+m_Q)\frac{1+\not{v}}{2}\Bigg[\Bigg(\frac{k^+}{\sqrt{2}}  +\frac{m_q}{2}\Bigg)\not{n}_+
\nonumber\\
&&+\Bigg(\frac{k^-}{\sqrt{2}}  +\frac{m_q}{2}\Bigg)\not{n}_- -k_{\perp}^{\mu}\gamma_{\mu}  \Bigg]\gamma^5\nonumber\\
&&-(E_q+m_q)\frac{1-\not{v}}{2} \Bigg[  \Bigg(\frac{k^+}{\sqrt{2}}-\frac{m_q}{2}\Bigg)\not{n}_+
\nonumber\\
 && +\Bigg(\frac{k^-}{\sqrt{2}}-\frac{m_q}{2}\Bigg)\not{n}_--k_{\perp}^{\mu}\gamma_{\mu}\Bigg]\gamma^5
  \Bigg\}_{\alpha\beta},\label{eqm}
\end{eqnarray}
where  $E_Q$ and $E_q$ are the energies of the heavy and light quarks respectively,
$v$ the four-speed of the $B$ meson, i.e., $p_B^\mu=m_B v^\mu$, $n_\pm^\mu$ are two light-like vectors with $n_\pm^\mu=(1,0,0,\mp 1)$, and
\begin{equation}
k^\pm=\frac{E_q\pm k^3}{\sqrt{2}},\;\;\; k_\perp^\mu=(0,k^1,k^2,0). \label{kpm}
\end{equation}
$K(\vec{k})$ is a function proportional to the $B$-meson wave function
\begin{equation}
K(\vec{k})=\frac{2N_B\Psi_0(\vec{k})}{\sqrt{E_qE_Q(E_q+m_q)(E_Q+m_Q)}} \label{wave-k},
\end{equation}
and $\Psi_0(\vec{k})$ is the wave function of $B$ meson in the rest-frame
\begin{equation}\label{psi0}
\Psi_0(\vec{k})=a_1 e^{a_2|\vec{k}|^2+a_3|\vec{k}|+a_4},
\end{equation}
with the parameters $a_i$ ($i=1,\cdots, 4$) obtained as \cite{SY2017}
\begin{eqnarray}
	&&a_1=4.55_{-0.30}^{+0.40}\,\mathrm{GeV}^{-3/2},\quad\;
	a_2=-0.39_{-0.20}^{+0.15}\,\mathrm{GeV}^{-2},\nonumber\\
	&& a_3=-1.55\pm 0.20\,\mathrm{GeV}^{-1},\quad   a_4=-1.10_{-0.05}^{+0.10}.
\end{eqnarray}

The momentum of pion is large in $B$ decays, therefore wave function can be defined in the light-cone coordinate, as \cite{bra1990,bal1999}
\begin{eqnarray}
\langle \pi(p_\pi)|\bar{q}(y)_\rho q'(0)_\delta |0\rangle &=&\int dx d^2k_{q\perp}e^{ix p_\pi\cdot y-y_{\perp}\cdot k_{q\perp}}\nonumber\\
&&\times\Phi^\pi_{\delta\rho},
\end{eqnarray}
and the wave function $\Phi^\pi_{\delta\rho}$ is
\begin{eqnarray}
\Phi^\pi_{\delta\rho}&=&\frac{if_\pi}{4}\Bigg\{\not{p}_\pi\gamma_5\phi_\pi(x,k_{q\perp})
  -\mu_\pi\gamma_5\Bigg(\phi^\pi_P(x,k_{q\perp})\nonumber\\
   && -\sigma_{\mu\nu}p_\pi^\mu y^\nu\frac{\phi^\pi_\sigma(x,k_{q\perp})}{6}\Bigg)\Bigg\}_{\delta\rho},
\end{eqnarray}
here $f_\pi$ is the decay constant of pion, $\mu_\pi$ the chiral parameter, $\phi_\pi$, $\phi^\pi_P$ and $\phi^\pi_\sigma$ are twist-2 and twist-3 distribution functions of pion, respectively. $\Phi^\pi_{\delta\rho}$ can be changed to momentum space \cite{bf2001,wy2002}
\begin{eqnarray}
\Phi^\pi_{\delta\rho}&=&\frac{if_\pi}{4}\Bigg\{\not{p}_\pi\gamma_5\phi_\pi(x,k_{q\perp})
  -\mu_\pi\gamma_5\Bigg(\phi^\pi_P(x,k_{q\perp})\nonumber\\
   && -i\sigma_{\mu\nu}\frac{p_\pi^\mu \bar{p}_\pi^\nu}{p_\pi\cdot \bar{p}_\pi} \frac{\phi'^\pi_\sigma(x,k_{q\perp})}{6}\nonumber\\
   &&+i  \sigma_{\mu\nu}p_\pi^\mu\frac{\phi^\pi_\sigma(x,k_{q\perp})}{6}\frac{\partial}{\partial k_{q\perp\nu}}\Bigg)\Bigg\}_{\delta\rho},
\end{eqnarray}
where $\bar{p}_\pi$ is a 4-momentum with its moving direction opposite to that of pion, and $\phi'^{\pi}_\sigma (x,k_{q\perp})=\partial \phi^\pi_\sigma (x,k_{q\perp})/\partial x$.

\begin{figure}[bth]
	\epsfig{file=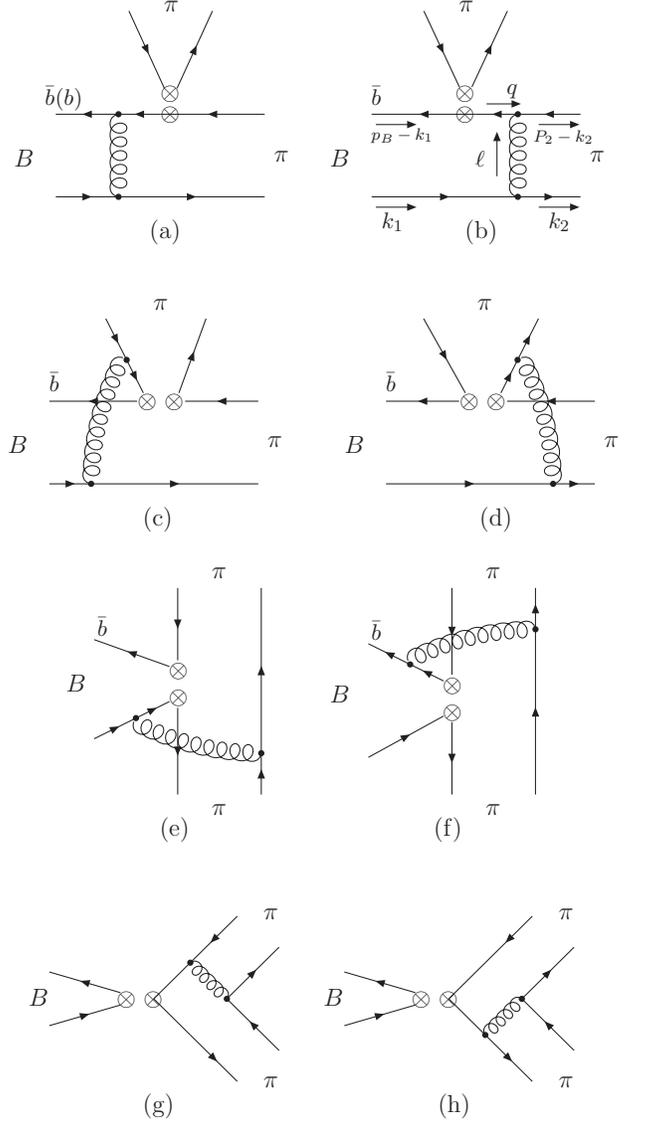,width=8cm,height=15.5cm} \caption{Diagrams contributing to the $B\rightarrow\pi\pi$ decays.} \label{fig1}
\end{figure}

The diagrams for the hard amplitude are shown in Fig. \ref{fig1}. Transverse momenta of quarks and gluons are kept in the calculation of the hard amplitude. Double logarithms such as $\alpha_s(\mu)\mbox{ln}^2 k_\text{T}/\mu$ in higher order radiative corrections in QCD are resummed into the Sudakov factor \cite{liyu1996-1,liyu1996-2}, and the double logarithms as $\alpha_s(\mu)\mbox{ln}^2 x$, here $x$ is the momentum fraction of partons in the longitudinal direction, can be resummed into the threshold factor \cite{lihn2002}. It is convenient for the $k_\text{T}$-resummation to be performed in $b$-space, with $b$ being the conjugate variable to $\bf{k}_T$. Therefore, it is relevantly convenient for the calculation of the hard amplitude to be also performed in the $b$-space.

The diagrams (a) and (b) in Fig. \ref{fig1} are factorizable diagrams. The amplitude contributed by these two diagrams with the insertion of operators of $(V-A)(V-A)$ current is

\begin{eqnarray}\label{fe}
&F_e &=-i\frac{4\pi^2}{N_c^2}f_B f_\pi^2 m_B\int dk_{\perp}k_{\perp}\int_{x^d}^{x^u}dx\int_0^1 dx_1  \nonumber\\
&&\int_0^\infty bdbb_1db_1 (\frac{1}{2}m_B+\frac{|\vec{k}_{\perp}|^2}{2x^2m_B})K(\vec{k})(E_Q+m_Q)\nonumber\\
&&\times J_0(k_{\perp}b) \Bigg\{\alpha_s(\mu_{e}^1)\Bigg(2m_B[E_q(1+x_1)+k^3(1-x_1)]\nonumber\\
&&\times \phi_\pi(x_1,b_1)+2\mu_\pi[E_q(1-2x_1)-k^3]\phi^\pi_P(x_1,b_1)\nonumber\\
&&+\frac{1}{3}\mu_\pi[E_q(1-2 x_1)-k^3]\phi'^\pi_\sigma(x_1,b_1)\Bigg)\nonumber\\
&&\times h_e^1(x,x_1,b,b_1)S_t(x_1)\exp[-S_{B}(\mu_{e}^1)-S_{\pi_1}(\mu_{e}^1)] \nonumber\\
&&+\alpha_s(\mu_{e}^2) [4\mu_\pi(E_q-k^3)]\phi^\pi_P(x_1,b_1)h_e^2(x,x_1,b,b_1)\nonumber\\
&& \times S_t(x)\exp[-S_{B}(\mu_{e}^2)-S_{\pi_1}(\mu_{e}^2)]\Bigg\}.
\end{eqnarray}

The contribution by the operators of $(S+P)(S-P)$ current which comes from the Fierz transformation of operators of $(V-A)(V+A)$ current is
\begin{eqnarray}\label{feP}
&F_e^P &=-i\frac{4\pi^2}{N_c^2}f_B f_\pi^2 \mu_\pi\int dk_{\perp}k_{\perp}\int_{x^d}^{x^u}dx\int_0^1 dx_1 \nonumber\\
&&\int_0^\infty bdbb_1db_1 (\frac{1}{2}m_B+\frac{|\vec{k}_{\perp}|^2}{2x^2m_B})K(\vec{k})(E_Q+m_Q)\nonumber\\
&&\times J_0(k_{\perp}b) \Bigg\{\alpha_s(\mu_{e}^1)\Bigg(4m_B(E_q+k^3)\phi_\pi(x_1,b_1)\nonumber\\
&&+4\mu_\pi[E_q(x_1+2)-k^3 x_1]\phi^\pi_P(x_1,b_1)\nonumber\\
&&+\frac{2}{3}\mu_\pi[k^3(x_1-2)-E_q x_1]\phi'^\pi_\sigma(x_1,b_1)\Bigg)\nonumber\\
&&\times h_e^1(x,x_1,b,b_1) S_t(x_1)\exp[-S_{B}(\mu_{e}^1)-S_{\pi_1}(\mu_{e}^1)] \nonumber\\
&&+\alpha_s(\mu_{e}^2) [ 8\mu_\pi(E_q-k^3) ]\phi^\pi_P(x_1,b_1)h_e^2(x,x_1,b,b_1)\nonumber\\
&& \times S_t(x)\exp[-S_{B}(\mu_{e}^2)-S_{\pi_1}(\mu_{e}^2)]\Bigg\}.
\end{eqnarray}
The contributions of diagrams of Fig. \ref{fig1} (c) and (d) are
\begin{eqnarray}
&M_e &=-i\frac{4\pi^2}{N_c^2}f_B f_\pi^2 m_B\int dk_{\perp}k_{\perp}\int_{x^d}^{x^u}dx\int_0^1 dx_1 dx_2 \nonumber\\
&&\int_0^\infty bdbb_2db_2 (\frac{1}{2}m_B+\frac{|\vec{k}_{\perp}|^2}{2x^2m_B})K(\vec{k})(E_Q+m_Q)\nonumber\\
&&\times J_0(k_{\perp}b)\phi_\pi(x_2,b_2) \Bigg\{\alpha_s(\mu_{d}^1)
\Bigg(-2m_B(x_2-1)\nonumber\\
&&\cdot (E_q+k^3)\phi_\pi(x_1,b)-2\mu_\pi x_1(E_q-k^3)\phi^\pi_P(x_1,b)\nonumber
\end{eqnarray}
\begin{eqnarray}\label{Me}
&&+\frac{1}{3}\mu_\pi x_1(E_q-k^3)\phi'^\pi_\sigma(x_1,b)\Bigg) h_d^1(x,x_1,x_2,b,b_2)\nonumber\\
&&\times S_t(x_1)\exp[-S_{B}(\mu_{d}^1)-S_{\pi_1}(\mu_{d}^1)-S_{\pi_2}(\mu_{d}^1)] \nonumber\\
&&+\alpha_s(\mu_{d}^2) \Bigg(-2m_B[E_q(x_1+x_2)+k^3(x_2-x_1)]\nonumber\\
&&\cdot \phi_\pi(x_1,b)+2\mu_\pi x_1(E_q+k^3)\phi^\pi_P(x_1,b)\nonumber\\
&&+\frac{1}{3}\mu_\pi x_1(E_q+k^3)\phi'^\pi_\sigma(x_1,b)\Bigg)\nonumber h_d^2(x,x_1,x_2,b,b_2)\nonumber\\
&&\times S_t(x_1)\exp[-S_{B}(\mu_{d}^2)-S_{\pi_1}(\mu_{d}^2)-S_{\pi_2}(\mu_{d}^2)]\Bigg\}
\end{eqnarray}
for the operators of $(V-A)(V-A)$ current, and
\begin{eqnarray}\label{MeP}
&M_e^P &=-i\frac{4\pi^2}{N_c^2}f_B f_\pi^2 m_B\int dk_{\perp}k_{\perp}\int_{x^d}^{x^u}dx\int_0^1 dx_1 dx_2 \nonumber\\
&&\int_0^\infty bdbb_2db_2 (\frac{1}{2}m_B+\frac{|\vec{k}_{\perp}|^2}{2x^2m_B})K(\vec{k})(E_Q+m_Q)\nonumber\\
&&\times J_0(k_{\perp}b)\phi_\pi(x_2,b_2) \Bigg\{\alpha_s(\mu_{d}^1)
\Bigg(-2m_B(E_q(x_1-x_2\nonumber\\
&&+1)-k^3(x_1+x_2-1))\phi_\pi(x_1,b)\nonumber\\
&&+2\mu_\pi x_1(E_q+k^3)\phi^\pi_P(x_1,b)+\frac{1}{3}\mu_\pi x_1(E_q+k^3)\nonumber\\
&&\cdot \phi'^\pi_\sigma(x_1,b)\Bigg) h_d^1(x,x_1,x_2,b,b_2)\nonumber\\
&&\times S_t(x_1)\exp[-S_{B}(\mu_{d}^1)-S_{\pi_1}(\mu_{d}^1)-S_{\pi_2}(\mu_{d}^1)] \nonumber\\
&&+\alpha_s(\mu_{d}^2) \Bigg(2m_Bx_2(E_q+k^3)\phi_\pi(x_1,b)\nonumber\\
&&-2\mu_\pi x_1(E_q-k^3)\phi^\pi_P(x_1,b)+\frac{1}{3}\mu_\pi x_1\nonumber\\
&&\cdot(E_q-k^3)\phi'^\pi_\sigma(x_1,b)\Bigg) h_d^2(x,x_1,x_2,b,b_2)\nonumber\\
&&\times S_t(x_1)\exp[-S_{B}(\mu_{d}^2)-S_{\pi_1}(\mu_{d}^2)-S_{\pi_2}(\mu_{d}^2)]\Bigg\}
\end{eqnarray}
for the insertion of Fierz transformed operators of $(S+P)(S-P)$ current. The  contribution of operators of $(V-A)(V+A)$ current always vanishes.

The contributions of Fig. \ref{fig1} (e) and (f) are

\begin{eqnarray}\label{Ma}
M_a &=&-i\frac{4\pi^2}{N_c^2 }f_B f_\pi^2 \int dk_{\perp}k_{\perp}\int_{x^d}^{x^u}dx\int_0^1 dx_1 dx_2 \nonumber\\
&&\int_0^\infty bdbb_1db_1 (\frac{1}{2}m_B+\frac{|\vec{k}_{\perp}|^2}{2x^2m_B})K(\vec{k}) (E_Q+m_Q)\nonumber\\
&&\times J_0(k_{\perp}b)\Bigg\{\alpha_s(\mu_{f}^1)
\Bigg[-2m_B^2(x_2-1)(E_q+k^3) \nonumber\\
&&\cdot \phi_\pi(x_1,b_1)\phi_\pi(x_2,b_1)+\frac{1}{3}\mu_\pi^2\phi^\pi_P(x_1,b_1)\nonumber \\
&&\cdot\Big([E_q(x_1+x_2-1)-k^3(x_1-x_2+1)]\phi'^\pi_\sigma(x_2,b_1)\nonumber\\
&& +6[E_q(x_1-x_2+1)-k^3(x_1+x_2-1)]\phi^\pi_P(x_2,b_1)\Big)\nonumber\\
&& +\frac{1}{18}\mu_\pi^2\phi'^\pi_\sigma(x_1,b_1)\Big(-[E_q(x_1-x_2+1)-k^3(x_1 \nonumber\\
&&+x_2-1)]\phi'^\pi_\sigma(x_2,b_1)+6[E_q(x_1+x_2-1)+k^3(-x_1\nonumber\\
&&+x_2-1)]\phi^\pi_P(x_2,b_1) \Big)\Bigg]h_f^1(x,x_1,x_2,b,b_1)\nonumber\\
&&\times S_t(x_1)S_t(x_2)\exp[-S_{B}(\mu_{f}^1)-S_{\pi_1}(\mu_{f}^1)-S_{\pi_2}(\mu_{f}^1)]
 \nonumber\\
&&+\alpha_s(\mu_{f}^2) \Bigg[-2m_B^2x_1(E_q-k^3)\phi_\pi(x_1,b_1)\phi_\pi(x_2,b_1)\nonumber\\
&& -\frac{1}{3}\mu_\pi^2\phi^\pi_P(x_1,b_1) \Big(-[E_q(x_1+x_2-1)+k^3(-x_1\nonumber\\
&&+x_2+1)]\phi'^\pi_\sigma(x_2,b_1)+6[E_q(x_1-x_2+3)-k^3(x_1\nonumber\\
&&+x_2-1)]\phi^\pi_P(x_2,b_1)\Big) -\frac{1}{18}\mu_\pi^2\phi'^\pi_\sigma(x_1,b_1) \nonumber\\
&&\cdot\Big(-[E_q(x_1-x_2-1)-k^3(x_1+x_2-1)]\phi'^\pi_\sigma(x_2,b_1)\nonumber\\
&& +6[E_q(x_1+x_2-1)+k^3(-x_1+x_2-3)]\phi^\pi_P(x_2,b_1) \Big)\Bigg]\nonumber\\
&& \times h_f^2(x,x_1,x_2,b,b_1)\exp[-S_{B}(\mu_{f}^2)-S_{\pi_1}(\mu_{f}^2)\nonumber\\
&&-S_{\pi_2}(\mu_{f}^2)]\Bigg\}
\end{eqnarray}
for $(V-A)(V-A)$ current, and
\begin{eqnarray}\label{MaP}
&M_a^P &=-i\frac{4\pi^2}{N_c^2 }f_B f_\pi^2 \int dk_{\perp}k_{\perp}\int_{x^d}^{x^u}dx\int_0^1 dx_1 dx_2 \nonumber\\
&&\int_0^\infty bdbb_1db_1 (\frac{1}{2}m_B+\frac{|\vec{k}_{\perp}|^2}{2x^2m_B})K(\vec{k})(E_Q+m_Q)\nonumber\\
&&\times J_0(k_{\perp}b) \Bigg\{\alpha_s(\mu_{f}^1)\Bigg[2m_B^2x_1(E_q-k^3)\phi_\pi(x_1,b_1)\nonumber\\
&&\cdot\phi_\pi(x_2,b_1)+\frac{1}{3}\mu_\pi^2\phi^\pi_P(x_1,b_1)\Big(-[E_q(x_1+x_2-1)\nonumber\\
&&+k^3(-x_1+x_2-1)]\phi'^\pi_\sigma(x_2,b_1)+6[E_q(x_1-x_2+1)\nonumber\\
&&-k^3(x_1+x_2-1)]\phi^\pi_P(x_2,b_1)\Big) +\frac{1}{18}\mu_\pi^2\phi'^\pi_\sigma(x_1,b_1) \nonumber
\end{eqnarray}
\begin{eqnarray}
&&\cdot\Big(-[E_q(x_1-x_2+1)-k^3(x_1+x_2-1)]\phi'^\pi_\sigma(x_2,b_1)\nonumber\\
&& +6[E_q(x_1+x_2-1)+k^3(-x_1+x_2-1)]\phi^\pi_P(x_2,b_1) \Big)\Bigg]\nonumber\\
&&\times h_f^1(x,x_1,x_2,b,b_1)S_t(x_1)S_t(x_2)\exp[-S_{B}(\mu_{f}^1)\nonumber\\
&&-S_{\pi_1}(\mu_{f}^1)-S_{\pi_2}(\mu_{f}^1)] +\alpha_s(\mu_{f}^2) \Bigg[-2m_B^2(x_2-1)\nonumber \\
&&\cdot(E_q+k^3)\phi_\pi(x_1,b_1)\phi_\pi(x_2,b_1) +\frac{1}{3}\mu_\pi^2\phi^\pi_P(x_1,b_1) \nonumber\\
&&\Big( -[E_q(x_1+x_2-1)+k^3(-x_1+x_2-3)]\phi'^\pi_\sigma(x_2,b_1)\nonumber\\
&& -6[E_q(x_1-x_2+3)-k^3(x_1+x_2-1)]\phi^\pi_P(x_2,b_1)\Big)\nonumber\\
&& +\frac{1}{18}\mu_\pi^2\phi'^\pi_\sigma(x_1,b_1) \Big([E_q(x_1-x_2-1)-k^3(x_1+x_2\nonumber\\
&&-1)]\phi'^\pi_\sigma(x_2,b_1) +6[E_q(x_1+x_2-1)+k^3(-x_1+x_2\nonumber\\
&&+1)]\phi^\pi_P(x_2,b_1) \Big)\Bigg] h_f^2(x,x_1,x_2,b,b_1)\exp[-S_{B}(\mu_{f}^2)\nonumber\\
&&-S_{\pi_1}(\mu_{f}^2)-S_{\pi_2}(\mu_{f}^2)]\Bigg\}
\end{eqnarray}
for $(S+P)(S-P)$ current,
\begin{eqnarray}\label{MaR}
&M_a^R &=-i\frac{4\pi^2}{N_c^2}f_B f_\pi^2 m_B\mu_\pi\int dk_{\perp}k_{\perp}\int_{x^d}^{x^u}dx\int_0^1 dx_1 dx_2 \nonumber\\
&&\int_0^\infty bdbb_1db_1 (\frac{1}{2}m_B+\frac{|\vec{k}_{\perp}|^2}{2x^2m_B})K(\vec{k})(E_Q+m_Q)\nonumber\\
&&\times J_0(k_{\perp}b) \Bigg\{\alpha_s(\mu_{f}^1)\Bigg(-\frac{1}{3}(E_q+k^3)\nonumber\\
&&\cdot\phi_\pi(x_1,b_1)\Big((1-x_1)\phi'^\pi_\sigma(x_2,b_1) +6(x_1-1)\nonumber\\
&&\cdot\phi^\pi_P(x_2,b_1)\Big)-2 x_2(E_q-k^3)\phi_\pi(x_2,b_1)\phi^\pi_P(x_1,b_1)\nonumber\\
&&-\frac{1}{3} x_2(E_q-k^3)\phi_\pi(x_2,b_1)\phi'^\pi_\sigma(x_1,b_1)\Bigg)\nonumber\\
&&\times h_f^1(x,x_1,x_2,b,b_1)S_t(x_1)S_t(x_2)\exp[-S_{B}(\mu_{f}^1)\nonumber\\
&&-S_{\pi_1}(\mu_{f}^1)-S_{\pi_2}(\mu_{f}^1)] +\alpha_s(\mu_{f}^2) \Bigg(\frac{1}{3}\Big((E_q(x_1-2)\nonumber\\
&&-k^3x_1)\phi'^\pi_\sigma(x_2,b_1) -6(E_q(x_1-2)-k^3 x_1)\nonumber\\
&&\cdot\phi^\pi_P(x_2,b_1)\Big)\phi_\pi(x_1,b_1)-2 (E_q(x_2+1)+k^3(x_2 \nonumber\\
&&-1))\phi_\pi(x_2,b_1)\phi^\pi_P(x_1,b_1)-\frac{1}{3} (E_q(x_2+1)\nonumber\\
&&+k^3(x_2-1))\phi_\pi(x_2,b_1)\phi'^\pi_\sigma(x_1,b_1)\Bigg) h_f^2(x,x_1,x_2,b,b_1)\nonumber\\
&& \times\exp[-S_{B}(\mu_{f}^2)-S_{\pi_1}(\mu_{f}^2)-S_{\pi_2}(\mu_{f}^2)]\Bigg\}
\end{eqnarray}
for $(V-A)(V+A)$ current.

For the diagrams (g) and (h) in Fig. (\ref{fig1}), the contributions of operators of $(V-A)(V-A)$ are always cancel each other. Only the current of $(S+P)(S-P)$ from the Fierz transformation of operator of $(V-A)(V+A)$ current contributes. The result is
\begin{eqnarray}
&F_a^P &=-i\frac{8\pi  }{N_c^2}\chi_B f_\pi^2 \mu_\pi \int_0^1 dx_1 dx_2 \int_0^\infty b_1db_1b_2db_2 \nonumber\\
&&\times\Bigg\{\alpha_s(\mu_{a}^1)\Bigg(-4\phi^\pi_P(x_1,b_1)\phi_\pi(x_2,b_2)\nonumber\\
&&-\frac{1}{3} \Bigg[ (1-x_2)\phi'^\pi_\sigma(x_2,b_2)-6(x_2-1)\phi^\pi_P(x_2,b_2)\Bigg]\nonumber\\
&&\cdot \phi_\pi(x_1,b_1)\Bigg)h_a^1(x_1,x_2,b_1,b_2)S_t(x_2) \exp[-S_{\pi_1}(\mu_{a}^1)\nonumber\\
&&-S_{\pi_2}(\mu_{a}^1)] +\alpha_s(\mu_{a}^2) \Bigg(-4\phi_\pi(x_1,b_1)\phi^\pi_P(x_2,b_2)\nonumber
\end{eqnarray}
\begin{eqnarray}\label{faP}
&&+ \Bigg[ -\frac{1}{3}x_1\phi'^\pi_\sigma(x_1,b_1)+2x_1\phi^\pi_P(x_1,b_1)\Bigg]\phi_\pi(x_2,b_2)\Bigg)\nonumber \\
&& \times h_a^2(x_1,x_2,b_1,b_2)S_t(x_1)\exp[-S_{\pi_1}(\mu_{a}^2)-S_{\pi_2}(\mu_{a}^2)]\Bigg\},\nonumber\\
\end{eqnarray}
where
\begin{eqnarray} \label{chiB}
\chi _B &&= \pi f_B m_B \int dk_{\perp}k_{\perp}\int_{x^d}^{x^u}dx(\frac{1}{2}m_B+\frac{|\vec{k}_{\perp}|^2}{2x^2m_B})   \nonumber\\
	&&\cdot K(\vec{k}) \Bigg[(E_q+m_q)(E_Q+m_Q)+(E_q^2-m_q^2)\Bigg].
\end{eqnarray}
In the above equations, i.e., Eqs. (\ref{fe})$\sim$(\ref{faP}), the exponentials $\exp[-S_{B}(\mu)]$, $\exp[-S_{\pi_1}(\mu)]$ and $\exp[-S_{\pi_2}(\mu)]$ are the Sudakov factors which are associated with each meson at the relevant energy scale, which are given in the Appendix \ref{a}. $\phi_\pi(x,b)$, $\phi^\pi_P(x,b)$, and $\phi^\pi_\sigma(x,b)$ are the wave functions of pion in $b$-space, which can be found in Appendix B of Ref. \cite{Lu-Yang2021}. The functions $h_i$'s are Fourior transformations of the hard amplitudes, which are
\begin{widetext}
\begin{eqnarray}
h_e^1(x,x_1,b,b_1)&=&K_0(\sqrt{xx_1}m_B b)\Big[\theta (b-b_1)I_0(\sqrt{x_1}m_B b_1)K_0(\sqrt{x_1}m_B b)
+\theta(b_1-b)\times I_0(\sqrt{x_1}m_B b)K_0(\sqrt{x_1}m_B b_1)\Big],\nonumber\\
&&
\end{eqnarray}
\begin{eqnarray}
h_e^2(x,x_1,b,b_1)&=&K_0(\sqrt{xx_1}m_B b)  \Big[\theta(b-b_1)I_0(\sqrt{x}m_B b_1)K_0(\sqrt{x}m_B b)+\theta(b_1-b) \times I_0(\sqrt{x}m_B b)K_0(\sqrt{x}m_B b_1)\Big],\nonumber\\
&&
\end{eqnarray}
\begin{eqnarray}
h_d^1(x,x_1,x_2,b,b_2)&=&K_0(-i\sqrt{x_1(1-x_2)}m_Bb_2)\Big[\theta(b_2-b)I_0(\sqrt{xx_1}m_B b)K_0(\sqrt{xx_1}m_Bb_2)+\theta(b-b_2) I_0(\sqrt{xx_1}m_Bb_2)\nonumber\\
&& \times K_0(\sqrt{xx_1}m_B b)\Big],
\end{eqnarray}
\begin{eqnarray}
h_d^2(x,x_1,x_2,b,b_2)&=&K_0(-i\sqrt{x_1x_2}m_B b_2)\Big[\theta(b_2-b)I_0(\sqrt{xx_1}m_B b)K_0(\sqrt{xx_1}m_B b_2)+\theta(b-b_2)I_0(\sqrt{xx_1}m_B b_2)\nonumber\\
&&\times K_0(\sqrt{xx_1}m_B b)\Big],
\end{eqnarray}
\begin{eqnarray}
h_f^1(x_1,x_2,b,b_1)&=&K_0(-i\sqrt{x_1(1-x_2)}m_B b)\Big[\theta(b-b_1)I_0(-i\sqrt{x_1(1-x_2)}m_B b_1)K_0(-i\sqrt{x_1(1-x_2)}m_B b)\nonumber\\
&&+\theta(b_1-b)I_0(-i\sqrt{x_1(1-x_2)}m_B b)K_0(-i\sqrt{x_1(1-x_2)}m_B b_1)\Big],
\end{eqnarray}
\begin{eqnarray}
h_f^2(x_1,x_2,b,b_1)&=&K_0(\sqrt{1-x_2+x_1x_2}m_Bb)\Big[\theta(b-b_1)I_0(-i\sqrt{x_1(1-x_2)}m_B b_1)K_0(-i\sqrt{x_1(1-x_2)}m_B b)\nonumber\\
&&+\theta(b_1-b)I_0(-i\sqrt{x_1(1-x_2)}m_B b)K_0(-i\sqrt{x_1(1-x_2)}m_B b_1)\Big],
\end{eqnarray}
\begin{eqnarray}
h_a^1(x_1,x_2,b_1,b_2)&=&K_0(-i\sqrt{x_1(1-x_2)}m_B b_1)\Big[\theta(b_2-b_1)I_0(-i\sqrt{1-x_2}m_B b_1)K_0(-i\sqrt{1-x_2}m_B b_2)\nonumber\\
&&+\theta(b_1-b_2)I_0(-i\sqrt{1-x_2}m_Bb_2)K_0(-i\sqrt{1-x_2}m_B b_1)\Big],
\end{eqnarray}
\begin{eqnarray}
h_a^2(x_1,x_2,b_1,b_2)&=&K_0(-i\sqrt{x_1(1-x_2)} b_1)\Big[\theta(b_2-b_1)I_0(-i\sqrt{x_1}m_B b_1)K_0(-i\sqrt{x_1}m_B b_2)+\theta(b_1-b_2)\nonumber\\
&&\cdot I_0(-i\sqrt{x_1}m_B b_2)K_0(-i\sqrt{x_1}m_B b_1)\Big].
\end{eqnarray}
\end{widetext}
In general the higher order radiative correction emerges as $\alpha_s(\mu) \mathrm{ln}(m/\mu)$, where $m$ denotes some mass scales in the physical process. Here the scale may be $m_B$, the longitudinal and transverse momenta of the intermediate quark and gluons in the decay process, etc. Therefore, it is beneficial to select $\mu$ as the largest mass scale appearing in the hard amplitude $H$, so that the largest logarithm in the higher order correction can be removed, i.e.
\begin{eqnarray}
\mu_e^1 &=& \max(\sqrt{x_1}m_B,\sqrt{xx_1}m_B,1/b,1/b_1),\nonumber\\
\mu_e^2 &=& \max(\sqrt{x}m_B,\sqrt{xx_1}m_B,1/b,1/b_1),\nonumber\\
\mu_d^1 &=& \max(\sqrt{xx_1}m_B,\sqrt{x_1(1-x_2)}m_B,1/b_1,1/b_2),\nonumber\\
\mu_d^2 &=& \max(\sqrt{xx_1}m_B,\sqrt{x_1x_2}m_B ,1/b_1,1/b_2),\nonumber\\
\mu_f^1 &=& \max(\sqrt{x_1(1-x_2)}m_B,1/b_1,1/b_2),\\
\mu_f^1 &=& \max(\sqrt{x_1(1-x_2)}m_B,\sqrt{1-x_2+x_1x_2}m_B),\nonumber\\
\mu_a^1 &=& \max(\sqrt{1-x_2}m_B,\sqrt{x_1(1-x_2)}m_B,1/b_1,1/b_2),\nonumber\\
\mu_a^2 &=& \max(\sqrt{x_1}m_B,\sqrt{x_1(1-x_2)},1/b_1,1/b_2).\nonumber
\end{eqnarray}

In terms of the matrix elements calculated according to the diagrams given in Fig. \ref{fig1}, i.e., Eqs. (\ref{fe})$\sim$ (\ref{faP}), the decay amplitudes of $B\to \pi\pi$ process are
\begin{eqnarray}\label{Mp+p-}
\mathcal{M}\bar{(B^0}&&\rightarrow\pi^+ \pi^-)\nonumber\\
=&&F_e \Big[ \xi_u(\frac{1}{3}C_1+C_2)-\xi_t(\frac{1}{3}C_3+C_4+\frac{1}{3}C_9+C_{10}) \Big] \nonumber\\
&&-F_e^P \xi_t\Big[ \frac{1}{3}C_5+C_6+\frac{1}{3}C_7+C_{8} \Big]+M_e\Big[ \xi_u(\frac{1}{3}C_1)\nonumber\\
&&-\xi_t(\frac{1}{3}C_3+\frac{1}{3}C_9) \Big]+M_a\Big[ \xi_u(\frac{1}{3}C_2)-\xi_t(\frac{1}{3}C_3  \nonumber\\
&&+\frac{2}{3}C_4-\frac{1}{6}C_9+\frac{1}{6}C_{10}) \Big] +M_a^R\Big[ -\xi_t(\frac{1}{3}C_5-\frac{1}{6}C_7) \Big]\nonumber\\
&&+M_a^P\Big[ -\xi_t(\frac{2}{3}C_6+\frac{1}{6}C_{8}) \Big]+F_a^P\Big[
 -\xi_t(\frac{1}{3}C_5\nonumber\\
&&+C_6-\frac{1}{6}C_7-\frac{1}{2}C_8) \Big],
\end{eqnarray}

\begin{eqnarray}\label{Mp0p0}
\sqrt{2}\mathcal{M}(\bar{B^0}&&\rightarrow\pi^0 \pi^0)\nonumber\\
=&&F_e \Big[ -\xi_u(C_1+\frac{1}{3}C_2)-\xi_t(\frac{1}{3}C_3+C_4+\frac{3}{2}C_7\nonumber\\
&&+\frac{1}{2}C_{8}-\frac{5}{3}C_9 -C_{10}) \Big]-F_e^P \xi_t\Big[ \frac{1}{3}C_5+C_6 \nonumber\\
&&-\frac{1}{6}C_7-\frac{1}{2}C_{8} \Big]+M_e\Big[ -\xi_u(\frac{1}{3}C_2)-\xi_t(\frac{1}{3}C_3\nonumber\\
&&-\frac{1}{6}C_9-\frac{1}{2}C_{10}) \Big] +M_e^P\Big[ -\xi_t(-\frac{1}{2}C_8) \Big] \nonumber\\
&&+M_a\Big[ \xi_u(\frac{1}{3}C_2)-\xi_t(\frac{1}{3}C_3+\frac{2}{3}C_4-\frac{1}{6}C_9 \nonumber\\
&& +\frac{1}{6}C_{10}) \Big]+M_a^R\Big[ -\xi_t(\frac{1}{3}C_5-\frac{1}{6}C_7) \Big]\nonumber\\
&&+M_a^P\Big[ -\xi_t(\frac{2}{3}C_6+\frac{1}{6}C_{8}) \Big] +F_a^P\Big[ -\xi_t(\frac{1}{3}C_5\nonumber\\
&&+C_6-\frac{1}{6}C_7-\frac{1}{2}C_8) \Big],
\end{eqnarray}
and
\begin{eqnarray}\label{Mp-p0}
\sqrt{2}\mathcal{M}(B^-&&\rightarrow\pi^- \pi^0)\nonumber\\
=&&F_e \Big[ \xi_u(\frac{4}{3}C_1+\frac{4}{3}C_2)-\xi_t(2C_9-\frac{3}{2}C_7-\frac{1}{2}C_8\nonumber\\
&&+2C_{10}) \Big] -F_e^P \xi_t\Big[ \frac{1}{2}C_7+\frac{3}{2}C_8 \Big]+M_e\Big[ \xi_u(\frac{1}{3}C_1\nonumber\\
&&+\frac{1}{3}C_2)-\xi_t(\frac{1}{2}C_9+\frac{1}{2}C_{10}) \Big]+M_e^P\Big[ -\xi_t(\frac{1}{2}C_{8}) \Big] \nonumber\\
&& 
\end{eqnarray}
where $\xi_u = V_{ub}V_{ud}^{*}$, $\xi_t=V_{tb} V_{td}^{*} $. The decay width is expressed as
\begin{equation}
	\Gamma(B \rightarrow f) =\frac{G_F^2 m_B^3}{128 \pi} |\mathcal{M}(B\rightarrow f)|^2.
\end{equation}
Sudakov factor can suppress the long-distance contribution in the decay amplitude \cite{liyu1996-1,liyu1996-2}. We reanalysed $B\to\pi$ transition form factor in Ref. \cite{Lu-Yang2021} with the wave function of $B$ meson derived from the QCD-inspired relativistic potential model \cite{Yang2012,LY2014,LY2015,SY2017,SY2019}, where the whole mass spectrum of $b$-flavored mesons is consistent with experimental data, and the transverse-momentum dependence in the $B$ wave function is automatically included. We find that the suppression effect to the soft dynamics of the Sudakov factor is dependent on the end-point behavior of $B$ wave function, and in the case of the new wave function we use, large part of soft contribution still left. To make perturbative calculation reliable, a soft momentum cutoff and soft form factors have to be introduced. In this work the cutoff scale is chosen as 1 GeV, which is in accord with the strong coupling constant in the range $\alpha_s/\pi <0.165$. The introduction of the soft form factor may change the factorization formula for $B$ decays, which will be discussed in Sect. \ref{softFM}

\subsection{Next-to-Leading-Order Corrections}

In this section, we calculate several most important next-to-leading-order (NLO) contributions to the $B \rightarrow \pi \pi$ decays that includes three parts: the vertex corrections, the quark loops, and the magnetic penguins, as what have been done in Ref. \cite{PQCD+NL}. The NLO corrections contribute to decay amplitudes by modifying the combinations of Wilson coefficients listed below

\begin{eqnarray}
&&	a_1(\mu) =C_2(\mu) +\frac{C_1(\mu)}{N_c}, \nonumber\\
&&	a_2(\mu) =C_1(\mu) +\frac{C_2(\mu)}{N_c}, \\
&&	a_i(\mu) =C_i(\mu) +\frac{C_{i\pm 1}(\mu)}{N_c}, ~~ i=3-10  \nonumber\\ \nonumber
\end{eqnarray}
where the upper (lower) sign applies when $i$ is odd (even).

\subsubsection{Vertex Correction}

The contributions of vertex corrections to the Wilson coefficients are \cite{QCDf1,QCDf2,QCDf3,PQCD+NL}
\begin{eqnarray}
&&	a_1(\mu) \rightarrow a_1(\mu)+ \frac{\alpha_s(\mu)}{4 \pi}C_F\frac{C_1(\mu)}{N_c}V_1 ,\nonumber\\
&&	a_2(\mu) \rightarrow a_2(\mu)+ \frac{\alpha_s(\mu)}{4 \pi}C_F\frac{C_2(\mu)}{N_c}V_2 ,\nonumber\\
&&	a_i(\mu) \rightarrow a_i(\mu)+ \frac{\alpha_s(\mu)}{4 \pi}C_F\frac{C_{i \pm 1}(\mu)}{N_c}V_i ,\space   i=3-10, \nonumber \\
\end{eqnarray}
The vertex function $V_i$ in the naive dimensional regularization (NDR) scheme are given by  \cite{QCDf1,QCDf2,QCDf3}
\begin{widetext}

\begin{equation}
	V_i=\left\{
\begin{aligned}
	&12\ln\frac{m_b}{\mu}-18+\int_{0}^{1}dx \phi_\pi(x)g(x),& \mathrm{for} \; i =1-4,9,10, \\
	-&12\ln\frac{m_b}{\mu} + 6 -\int_{0}^{1}dx \phi_\pi(x)g(1-x),& \mathrm{for} \; i =5,7,   \\
	-&6 +\int_{0}^{1}dx \phi_P^\pi(x)g(1-x),& \mathrm{for} \; i= 6,8
\end{aligned}
\right.
\end{equation}
where $\phi_\pi(x)$ and $\phi_P^\pi(x)$ are the twist-2 and -3 wave functions of the meson emitted from the weak vertex, respectively. The hard kernels are
\begin{equation}
	g(x) = 3 \left( \frac{1-2x}{1-x} \ln x -i \pi \right)+ \left[ 2\mathrm{Li}_2(x)- \ln^2 x-\frac{2 \ln x}{1-x} -(3+2i\pi)\ln x - (x \leftrightarrow 1 -x)  \right],
\end{equation}
\begin{equation}
	h(x) =   2\mathrm{Li}_2(x)- \ln^2 x -(1+2i\pi)\ln x - (x \leftrightarrow 1 -x)  .
\end{equation}

\end{widetext}
It has been shown that the inclusion of the vertex corrections can moderate the dependence of most of the Wilson coefficients on the renormalization scale $\mu$ \cite{QCDf1,PQCD+NL}.

\subsubsection{Contribution of Quark Loops}

For the $b\rightarrow d$ transition, the effective Hamiltonian contributed by the virtual quark loops can be given by \cite{PQCD+NL}
\begin{eqnarray}
	\mathcal{H}_{\mathrm{eff}} = &-&\sum_{q=u,c,t}\sum_{q'}\frac{G_F}{\sqrt{2}}V_{qb}V_{qd}^*\frac{\alpha_s(\mu)}{2\pi}C^{(q)}(\mu,l^2)  \nonumber \\
	&\times& (\bar{d}\gamma_{\rho}(1-\gamma_5)T^a b)(\bar{q}'\gamma^{\rho}T^a q'),
\end{eqnarray}
where the functions $C^{(q)}(\mu,l^2)$ can be written as
\begin{equation}\label{cq}
		C^{(q)}(\mu,l^2) =\left[G^{(q)}(\mu,l^2) -\frac{2}{3} \right]C_2(\mu)
\end{equation}
for $q =u,c$, and
\begin{eqnarray}\label{ct}
	C^{(t)}(\mu,l^2) =&&\left[  G^{(d)}(\mu,l^2) -\frac{2}{3}\right]C_3(\mu) \nonumber \\
		&+ & \sum_{q''=u,d,s,c}G^{(q'')}(\mu,l^2)\left[C_4(\mu) -C_6(\mu) \right]. \nonumber \\
\end{eqnarray}
The function $G$ in Eqs. (\ref{cq}) and (\ref{ct}) is
\begin{equation}
	 G^{(q)}(\mu,l^2) = -4 \int_{0}^{1}dx x(1-x) \ln \frac{m_q^2-x(1-x)l^2-i\varepsilon}{\mu^2},
\end{equation}
where $m_q$ is the mass of quark $q$ $(q=u,d,s,c)$. For $q=u$ and $d$, the quark mass can be taken to be $m_q=0$.

The quark-loop contribution can be absorbed into the Wilson Coefficients $a_4,a_6 $ because the topological structure of its contribution to the effective Hamiltonian is the same as the contribution of penguin diagram, then
\begin{equation}
	a_{4,6}(\mu) \rightarrow a_{4,6}(\mu)+\frac{\alpha_s(\mu)}{9\pi}\sum_{q=u,c,t}\frac{V_{qb}V_{	qd}^*}{V_{tb}V_{td}^*}C^{(q)}(\mu,\left<l^2\right>),
\end{equation}
where $\left<l^2\right>$ is the mean distribution of the momentum squared of the gluon that attached to the virtual quark loop and the final generated quark pair. One can take $\left<l^2\right>=m_b^2/4$ in the numerical analysis as a reasonable value in $B$ decays.

\subsubsection{Magnetic Penguins}

The effective Hamiltonian of magnetic penguin contains  the weak $ b \rightarrow d\textsl{g} $ transition
\begin{equation}
	\mathcal{H}_{\mathrm{eff}}= - \frac{G_F}{\sqrt{2}}V_{tb}V_{td}^*C_{8\textsl{g}}O_{8\textsl{g}},
\end{equation}
where the magnetic-penguin operator is given by
\begin{equation}
	O_{8\textsl{g}}=\frac{g}{8 \pi^2}m_b \bar{d}_i \sigma_{\mu \nu}(1+\gamma_5)T_{ij}^a G^{a\mu\nu} b_j.
\end{equation}
This Hamiltonian can contribute to the hadronic $B$ decay by the fission of the virtual gluon into a new quark-antiquark pair, which can be absorbed into the relevant Wilson coefficients \cite{PQCD+NL}
\begin{equation}
	a_{4,6}(\mu) \rightarrow a_{4,6}(\mu)-\frac{\alpha_s(\mu)}{9\pi} \frac{2m_B}{\sqrt{\left<l^2\right>}}C_{8\textsl{g}}^{\mathrm{eff}}(\mu),
\end{equation}
where the effective coefficient $C_{8\textsl{g}}^{\mathrm{eff}}=C_{8\textsl{g}}+C_5$ \cite{Hamiltanion1996}.
\vspace{0.5em}

\section{The contribution of the Soft Form Factors of $B\pi$ transition and $\pi\pi$ Production \label{softFM}}

We find that soft contributions in Eqs. (\ref{fe}), (\ref{feP}) and (\ref{faP}) which are relevant to the diagrams (a), (b), (g) and (h) in Fig. \ref{fig1} are still large in the case of the wave function of $B$ meson in Eq. (\ref{B-wave}) being used in this work, more than 40\% of the contribution is in the range of $\alpha_s/\pi >0.2$, while the contributions of the diagrams (c) and (d) in Fig. \ref{fig1} are dominated by perturbative contribution, more than 93\% of the contribution is in the range $\alpha_s/\pi <0.2$. For the diagrams (e) and (f) in Fig. \ref{fig1}, the contributions are only at the level of a few percent, which can be neglected. Therefore, to keep the perturbative calculation reliable, we introduce a momentum cutoff, i.e., taking a stringent perturbative requirement with the hard scale $\mu_{\mathrm{h}} >1.0~\mathrm{GeV}$, which is relevant to $\alpha_s/\pi <0.165$. The contributions lower than the hard scale with $\mu <1.0~\mathrm{GeV}$ are replaced by two kinds of soft form factors, the soft $B\pi$ transition form factor and $\pi\pi$ production form factor. Then the total $B\pi$ transition form factor is separated into two parts
\begin{equation} \label{softBpi}
F_0^{B\pi}=h_0^{B\pi}+\xi^{B\pi},
\end{equation}
where $h_0^{B\pi}$ is the $B\pi$ transition form factor that is dominated by the hard contribution, and $\xi^{B\pi}$ the soft transition form factor. The hard form factor can be calculated in the perturbation QCD approach which is relevant to the diagrams of Fig. \ref{fig1} (a) and (b), while the soft contribution of these two diagrams can be written in terms of the soft form factor $\xi^{B\pi}$. Including the contribution of the soft form factor, the amplitude is changed as
\begin{eqnarray}
	\mathcal{M} \rightarrow \mathcal{M}&-&2if_{\pi}C(\mu_{h})V_{\mathrm{CKM}} \cdot\xi^{B\pi}\nonumber\\
	&-&4i\frac{\mu_\pi}{m_B}f_{\pi}C'(\mu_{h})V_{\mathrm{CKM}} \cdot\xi^{B\pi},
\end{eqnarray}
where $C(\mu_{h})$ and $C'(\mu_{h})$ are the corresponding Wilson coefficients of the operators of $(V-A)(V-A)$ and $(S+P)(S-P)$ at $\mu_{h}=1.0~\mathrm{GeV}$, respectively, where the scale is taken as the critical scale that separates the hard and soft contribution.

There are also soft contributions from the factorizable diagrams (g) and (h) in Fig.\ref{fig1}, which can be absorbed into the soft production form factor of $\pi\pi$. The soft $\pi\pi$ production form factor can be defined from the scalar current
\begin{equation}
	\langle\pi\pi|S|0\rangle =-\dfrac{1}{2} F_{+}^{\pi\pi}(q^2)\mu_{\pi},
\end{equation}
where $\mu_{\pi}=m_\pi^2/(m_u+m_d)$ for the charged pion, which can be treated as a phenomenological parameter and taken as $\mu_{\pi}=1.75 ~\mathrm{GeV}$. The form factor $F_{+}^{\pi\pi}$ can be separated into two parts, the hard and soft parts
\begin{equation}
F_{+}^{\pi\pi}=h^{\pi\pi}+\xi^{\pi\pi},
\end{equation}
here $h^{\pi\pi}$ is the hard part of the $\pi\pi$ production form factor, which can be calculated in the perturbative QCD approach, and $\xi^{\pi\pi}$ being the soft part of the production form factor. Then the soft form factor contributes to the amplitude as
\begin{equation}
	\mathcal{M} \rightarrow \mathcal{M}+\frac{2\mu_{\pi}}{m_B^2} \langle0|S-P|B\rangle C(\mu_{h})V_{\mathrm{CKM}} \xi^{\pi\pi},
\end{equation}
where $\langle0|S-P|B\rangle=-i\chi_B$, and $\chi_B$ can be found in Eq. (\ref{chiB}).

\section{color-octet Matrix Element}

In this section, to explain the experimental data on the branching ratios and $CP$ violation we consider the possible contribution of the matrix element of operators composed of color-octet current. Consider a four-quark operator $(\bar{q}_{1i} q_{2j})(\bar{q}_{3j} q_{4i})$, where $i,~j$ are the color indices, and the current can be with any Dirac spinor structure. Due to the relation for the generators of the color SU(3) group
\begin{equation}
	T_{ik}^a T_{jl}^a = -\frac{1}{2N} \delta _{ik} \delta_{jl}+\frac{1}{2}\delta_{il}\delta_{jk},
\end{equation}
the four-quark operator can be transferred to color singlet and octet operators
\begin{equation}
	(\bar{q}_{1i} q_{2j})(\bar{q}_{3j} q_{4i})
		= \frac{1}{N} (\bar{q}_{1i} q_{2i})(\bar{q}_{3j} q_{4j})+2(\bar{q}_{1}T^a q_{2})(\bar{q}_{3}T^a q_{4}).
\end{equation}
As for the decay of $B\to \pi\pi$, we can take the contribution of the operators $(V-A)(V-A)$ and $(S+P)(S-P)$ to $\bar{B}^0\to \pi^+\pi^-$ decay as an example. We define
\begin{eqnarray}
&&T_{\pi\pi}^8=\langle \pi^-\pi^+|(\bar{d}T^a u)_{V-A} (\bar{u}T^a b)_{V-A}|\bar{B}^0\rangle ,\\
&&T_{\pi\pi}^{SP8}=\langle \pi^-\pi^+|(\bar{d}T^a u)_{S+P} (\bar{u}T^a b)_{S-P}|\bar{B}^0\rangle,
\end{eqnarray}
where a convention is indicated that the quarks in the first current flow into the first meson in the final state, and the quarks in the second current involve the initial and second meson of the final state. Such kinds of color-octet matrix elements have all been dropped previously because mesons in both the initial and final states should be in color-singlet. In this work we assume that such color-octet current can give nonzero contribution. The quark pair in color-octet can transfer into singlet state by exchanging soft gluons with other quark system at distance of hadronic scale. Certainly such color-octet matrix element should be smaller compared with color-singlet contribution. One can assume that the momentum and spin structure of the quark system will not change much when exchanging soft gluons.

The contribution of the color-octet matrix element to the amplitude $\mathcal{M}(\bar{B}^0\to \pi^+\pi^-)$ is
\begin{eqnarray}
\mathcal{M}_{\pi^+\pi^-}^8&=&\frac{2}{m_B^2}\bigg[V_{ub}V_{ud}^*\big(2C_1T_{\pi\pi}^8\big)
+\sum_{q=u,c}V_{qb}V_{qd}^*\cdot 2\big[(C_3\nonumber\\
&&+\frac{3}{2}e_u C_9)T_{\pi\pi}^8-(2C_5+3e_u C_7)T_{\pi\pi}^{SP8}\big]\bigg].
\end{eqnarray}
The above color-octet contribution should be added to the former amplitude $\mathcal{M}$ as
\begin{equation}
\mathcal{M}\to \mathcal{M}+\mathcal{M}_{\pi^+\pi^-}^8.
\end{equation}
Similarly the color-octet contribution to the decays of $B^-\to\pi^-\pi^0$ and $\bar{B}^0\to\pi^0\pi^0$ are
\begin{eqnarray}
\sqrt{2}\mathcal{M}_{\pi^-\pi^0}^8&=&\frac{2}{m_B^2}\bigg[V_{ub}V_{ud}^*\big(2(C_1+C_2)T_{\pi\pi}^8\big)
+\sum_{q=u,c}V_{qb}V_{qd}^*\nonumber\\
&&\cdot \frac{3}{2}(e_u-e_d)\big[2(-C_8+ C_9+C_{10})T_{\pi\pi}^8-4C_7\nonumber\\
&&\cdot T_{\pi\pi}^{SP8}\big]\bigg],
\end{eqnarray}
and
\begin{eqnarray}
\sqrt{2}\mathcal{M}_{\pi^0\pi^0}^8&=&\frac{2}{m_B^2}\bigg[-V_{ub}V_{ud}^*\big(2C_2T_{\pi\pi}^8\big)
-\sum_{q=u,c}V_{qb}V_{qd}^*\big[2\nonumber\\
&&\cdot[-C_3+\frac{3}{2}(e_u-e_d)(-C_8+C_{10})-\frac{3}{2}e_dC_9]T_{\pi\pi}^8\nonumber\\
&&+2(2C_5+3e_dC_7) T_{\pi\pi}^{SP8}\big]\bigg].
\end{eqnarray}
In the above two equations, the following relations have been used
\begin{equation}
\langle \pi^0\pi^0|(\bar{u}T^a u)_{V+A} (\bar{d}T^a b)_{V-A}|\bar{B}^0\rangle
\approx  \frac{1}{2}T_{\pi\pi}^8
\end{equation}
and
\begin{equation}
\langle \pi^0\pi^0|(\bar{d}T^a d)_{V+A} (\bar{d}T^a b)_{V-A}|\bar{B}^0\rangle
\approx  -\frac{1}{2}T_{\pi\pi}^8
\end{equation}
The color-octet contribution $\mathcal{M}_{\pi^-\pi^0}^8$ and $\mathcal{M}_{\pi^0\pi^0}^8$ should also be added to the former amplitudes $\mathcal{M}$ for each relevant decay mode.

To show the relative magnitude of the color-octet to singlet hadronic matrix element, we can define two parameters $\delta_8$ and $\delta_8^{SP}$. Since the color-singlet hadronic matrix elements are approximately in accord with the follow result
\begin{eqnarray}
	&&\langle \pi^-\pi^+|(\bar{d} u)_{V-A} (\bar{u} b)_{V-A}|\bar{B}^0\rangle \approx -if_{\pi}m_B^{2}F_0^{B\pi}(0),  \nonumber \\
	&&\langle \pi^-\pi^+|(\bar{d} u)_{S+P} (\bar{u} b)_{S-P}|\bar{B}^0\rangle  \approx  -if_{\pi}\mu_{\pi}m_BF_0^{B\pi}(0),\nonumber \\
\end{eqnarray}
under the naive factorization approximation. So one can define $\delta_8$ and $\delta_8^{SP}$ as
\begin{eqnarray}
	T_{\pi\pi}^8 &=& -if_{\pi}m_B^{2}F_0^{B\pi}(0)\delta_8,  \nonumber \\
	T_{\pi\pi}^{SP8}  &=& -if_{\pi}\mu_{\pi}m_BF_0^{B\pi}(0)\delta_8^{SP}.
\end{eqnarray}
The smaller the parameters $\delta_8$ and $\delta_8^{SP}$, the smaller the color-octet matrix element relative to color-singlet one.

\section{Numerical analysis and discussion}

The input parameters in the numerical calculation are the soft $B\pi$ transition form factor $\xi^{B\pi}$, the soft $\pi\pi$ production form factor $\xi^{\pi\pi}$, and the color-octet matrix element parameter $\delta_8$ and $\delta_8^{SP}$ except for the parameters in $B$ and pion wave functions.


The hard part of $B\pi$ transition form factor $h_0^{B\pi}$ is relevant to the diagrams (a) and (b) of Fig.\ref{fig1} except the decay constant of the emitted pion. The hard form factor can be calculated in perturbation method with the hard scale $\mu_h>1\mathrm{GeV}$, which corresponds to $\alpha_s/\pi<0.165$. The result is
\begin{equation}
h_0^{B\pi}=0.23\pm 0.01,
\end{equation}
while the total $B\pi$ transition form factor is
\begin{equation}
	F_0^{B\pi}(0)= F_+^{B\pi}(0) = 0.27\pm 0.02,
\end{equation}
which is consistent with the measured differential branching ratio of $\bar{B}^0\to\pi^+\ell\nu$ in the $q^2\sim 0$ region in  experiment \cite{belle2013}. Then according to Eq. (\ref{softBpi}), the soft part of $B\pi$ transition form factor is
\begin{equation}
	\xi^{B\pi} = 0.04\pm 0.01.
\end{equation}

For the color-octet parameters $\delta_8$ and $\delta_8^{SP}$, we assume they are at the same order, and simply take
\begin{equation}
	\delta_8=\delta_8^{SP}
\end{equation}
to decrease the number of free inputs. Then the remaining free input parameters are only $\xi^{\pi\pi}$ and $\delta_8$, which can be obtained by fitting the experimental data of the branching ratios and $CP$ violation for the three $B\to\pi\pi$ decay modes. $\xi^{\pi\pi}$ and $\delta_8$ can be set in the following form
\begin{eqnarray}
	\delta_8 = d_1 e^{i\phi_1 } ,  \nonumber\\
	\xi^{\pi\pi} = d_2 e^{i\phi_2 },
\end{eqnarray}
where $\phi_1$ and $\phi_2$ are the strong phases of the color-octet matrix element and the soft $\pi\pi$ production form factor, respectively. Since $\pi\pi$ production form factor is time-like, it may have nonzero phase. The parameters $d_1$, $\phi_1$, $d_2$ and $\phi_2$ are fitted to the experimental data. The values that can reproduce all the branching ratios and $CP$ violation consistent with experimental data are found to be
\begin{eqnarray}
	&&d_1 = 0.250\pm0.015,~~~\phi_1 = (-0.440\pm0.016)\pi, \nonumber\\
	&&d_2 = 0.17\pm0.02,~~~~~~\phi_2 = (-0.76\pm0.03)\pi,
\end{eqnarray}
where the uncertainties mainly come from the constraint of the experimental data. The value of $d_1$ shows that the magnitude of the color-octet matrix element is only about 1/4 of the color-singlet contribution.

The branching ratios and $CP$ violation obtained in this work are
\begin{eqnarray}
	B(B^0\rightarrow\pi^+ \pi^-)&=&5.14\pm0.61^{+0.17+0.29}_{-0.28-0.24} \times 10^{-6}, \nonumber\\
	B(B^0\rightarrow\pi^0 \pi^0)&=&1.50\pm 0.24^{+0.04+0.18}_{-0.06-0.18} \times 10^{-6}, \nonumber\\
	B(B^+\rightarrow\pi^+ \pi^0)&=&5.72\pm0.44^{+0.19+0.22}_{-0.31-0.21} \times 10^{-6}, \nonumber\\
	A_{CP}(B^0\rightarrow\pi^+ \pi^-)&=&0.33 \pm0.04^{+0.01+0.04}_{-0.00-0.03},   \\
	A_{CP}(B^0\rightarrow\pi^0 \pi^0)&=&0.23\pm0.07^{+0.00+0.07}_{-0.00-0.05},    \nonumber\\
	A_{CP}(B^+\rightarrow\pi^+ \pi^0)&=&0.0054\pm0.0004^{+0.0000+0.0001}_{-0.0000-0.0001},\nonumber
\end{eqnarray}
where the uncertainties come from the variation of the theoretical input parameters. The first one is caused by the uncertainty of the soft parameters constrained by experiments, the second and last are caused by the variation of the parameters in B meson and pion wave functions, respectively. For the decay mode $B^+\rightarrow\pi^+ \pi^0$, the uncertainty of the $CP$ violation caused by the variation of the parameters in $B$ and pion wave functions are very small, which can be neglected.

The contributions of each theoretical component and the comparison of the total results with experimental data are presented in Table \ref{tbr-cp}.
\begin{widetext}
\begin{center}
	\begin{table*}[htb]
		\renewcommand\arraystretch{1.5}
		\caption{\label{tbr-cp} Branching ratios and direct CP violation ($\delta_8=\delta_8^{SP}$, $l^2 = \frac{m_b^2}{4}$, $m_c =1.3~\mathrm{GeV}$), where NLO is the hard contribution up to next-to-leading order in QCD, ``$+\xi_{B\pi}$" contribution of NLO + the contribution of the soft transition form factor $\xi_{B\pi}$, ``$+T_8$" contribution of NLO + color-octet matrix element, ``$+\xi_{\pi\pi}$" contribution of NLO + contribution of soft production form factor of $\pi\pi$, ``$+\xi_{B\pi}+T_8+\xi_{\pi\pi}$" total contribution of NLO$+\xi_{B\pi}+T_8+\xi_{\pi\pi}$, for which the first uncertainty comes from the constraint of experimental data, the second is the quadratic combination of uncertainties from the variation of input parameters in $B$ and pion wave functions. The last column is the experimental data from PDG \cite{PDG2020}.}
		\setlength{\tabcolsep}{4.2mm}{
			\begin{tabular}{c|c|c|c|c|c|c}
				\hline
				\hline
				Mode	& NLO	& $+\xi_{B\pi}$  & $+\xi_{\pi\pi}$ & $+T_8$ & $+\xi_{B\pi}+\xi_{\pi\pi}+T_8$  &Data \cite{PDG2020}\\ \hline
				B($B^0\rightarrow\pi^+ \pi^-$)$\times 10^{-6}$ &4.95 & 7.48   & 3.32  & 4.37 & $5.14\pm0.61^{+0.34}_{-0.37}$ & $5.12 \pm 0.19$   \\
				B($B^+\rightarrow\pi^+ \pi^0$)$\times 10^{-6}$    &3.27 & 4.40 & 3.27 &4.23 & $5.72\pm0.44^{+0.29}_{-0.37}$ &$5.5 \pm 0.4$    \\
				B($B^0\rightarrow\pi^0 \pi^0$)$\times 10^{-6}$ &0.13 & 0.14 & 0.22 & 0.67  & $1.50\pm0.24^{+0.18}_{-0.19}$ & $1.59 \pm 0.26$ \\
				$A_{CP}$($B^0\rightarrow\pi^+ \pi^-$) &0.17 & 0.11& 0.44 & 0.22 & $0.33\pm0.04^{+0.04}_{-0.03}$  & $0.32 \pm 0.04$  \\
				$A_{CP}$($B^+\rightarrow\pi^+ \pi^0$)   &-0.0007 & -0.0007& -0.0007 & 0.0053  & $0.0054\pm0.0004^{+0.0001}_{-0.0001}$ & $0.03 \pm 0.04$   \\
				$A_{CP}$($B^0\rightarrow\pi^0 \pi^0$)&0.27 & 0.48 & -0.16 & 0.53 & $0.23\pm0.07^{+0.07}_{-0.05}$  & $0.33 \pm 0.22$ \\
				\hline
				\hline
		\end{tabular}   }
	\end{table*}
\end{center}
\end{widetext}

It is shown in Table \ref{tbr-cp} that the soft transition form factor $\xi^{B\pi}$ can increase the amplitudes of $B^0\to\pi^+ \pi^-$, $B^+\to\pi^+ \pi^0$ and $B^0\to\pi^0 \pi^0$ by 23\%, 16\% and 4\%, respectively. The influence of $\xi^{B\pi}$ on $CP$ violation is tiny. The influence of $\xi^{\pi\pi}$ on branching ratios is generally smaller than $\xi^{B\pi}$, but it can increase the branching ratio of $B^0\to\pi^0 \pi^0$, which is the correct tendency to explain the large branching ratio of this decay mode measured in experiment. The contributions of the color-octet matrix element $T_8$ in  $B^0\to\pi^+ \pi^-$ and $B^+\to\pi^+ \pi^0$ are similar to $\xi^{\pi\pi}$, but it can increase the branching ratio of $B^0\to\pi^0 \pi^0$ more greatly, which is the key point to solve the $\pi\pi$ puzzle in our method. The final results for the branching ratios and $CP$ violations are given in the sixth column of Table \ref{tbr-cp}. The experimental data are also presented in the last column for comparison. It can be seen that all the branching ratios and $CP$ violation for $B\to \pi\pi$ decays are in good agreement with experimental data.

We note that the experimental data on $CP$ violation for $B^+\to\pi^+\pi^0$ and $B^0\to \pi^0\pi^0$ are still not so good in precision at present. More precise data on $A_{CP}$($B^+\rightarrow\pi^+ \pi^0$) and $A_{CP}$($B^0\rightarrow\pi^0 \pi^0$) in experiment are welcome, which may put more stringent constraint on our theoretical predictions.

A small comment would like to be given at the end of this section. The branching ratio of  $B^0\to \pi^0\rho^0$ predicted in PQCD \cite{PQCD4,luyang2002} is also awfully smaller than experimental data given in PDG \cite{PDG2020}. We believe that this problem could also be solved by the method suggested in this work. The none-zero color-octet hadronic matrix element relevant to $B\to\rho\pi$ decay can enhance the small branching ratio of $B^0\to \pi^0\rho^0$ without affecting the theoretical predictions for the other decay modes much as what happened in $B\to \pi\pi$ decays in this work. Such a research will be initiated soon in the near future.

\section{Summary}

We study $B\to\pi\pi$ decays in a modified perturbative QCD approach in this work. By using the wave function of $B$ meson obtained by solving the bound-state equation in relativistic potential model, the soft contribution to the decay processes can not be suppressed effectively by Sudakov factor. A soft momentum cutoff has to be introduced, and soft contributions are replaced by soft form factors. The nonzero color-octet hadronic matrix element is also introduced, which can enhance the usual color-suppressed contributions to the branching ratio of $B^0\to \pi^0\pi^0$ decay without affecting the branching ratio of the other two decay modes of $B^0\to\pi^+ \pi^-$ and $B^+\to\pi^+ \pi^0$ too much. By selecting the appropriate parameter space, all the branching ratios and $CP$ violations can be obtained in a good agreement with experimental data.

\vspace{0.5cm}
\acknowledgments
This work is supported in part by the National Natural Science Foundation of China under
Contracts No. 11875168.

\appendix{}
\section{\label{a}Sudakov factor and single ultraviolet logarithms in QCD}

The exponentials $\exp[-S_{B}(\mu)]$,  $\exp[-S_{\pi_1}(\mu)]$ and $\exp[-S_{\pi_2}(\mu)]$ are the combination of the Sudakov factor and the single ultraviolet logarithms associated with $B$ meson and pions. The exponents are defined as
\begin{equation}
S_B(\mu) = s(x,b,m_B)-\frac{1}{\beta_1}\ln \frac{\ln (\mu/\Lambda_{\mbox{QCD}})}
         {\ln (1/(b\Lambda_{\mbox{QCD}}))}
\end{equation}
\begin{eqnarray}
S_{\pi_1}(\mu) &=& s(x_1,b_1,m_B)+s(1-x_1,b_1,m_B)\nonumber\\
&&\;\;-\frac{1}{\beta_1}\ln \frac{\ln (\mu/\Lambda_{\mbox{QCD}})}
         {\ln (1/(b_1\Lambda_{\mbox{QCD}}))}
\end{eqnarray}
\begin{eqnarray}
S_{\pi_2}(\mu) &=& s(x_2,b_2,m_B)+s(1-x_2,b_2,m_B)\nonumber\\
&&\;\;-\frac{1}{\beta_1}\ln \frac{\ln (\mu/\Lambda_{\mbox{QCD}})}
         {\ln (1/(b_2\Lambda_{\mbox{QCD}}))}
\end{eqnarray}
The exponent $S(x,b,Q)$ up to next-leading order in QCD is \cite{Li1995}
\begin{widetext}
\begin{eqnarray}
&& s(x,b,Q)=\frac{A^{(1)}}{2\beta_{1}}\hat{q}\ln\left(\frac{\hat{q}}
{\hat{b}}\right)-
\frac{A^{(1)}}{2\beta_{1}}\left(\hat{q}-\hat{b}\right)+
\frac{A^{(2)}}{4\beta_{1}^{2}}\left(\frac{\hat{q}}{\hat{b}}-1\right)
-\left[\frac{A^{(2)}}{4\beta_{1}^{2}}-\frac{A^{(1)}}{4\beta_{1}}
\ln\left(\frac{e^{2\gamma_E-1}}{2}\right)\right]
\ln\left(\frac{\hat{q}}{\hat{b}}\right)
\nonumber \\
&&+\frac{A^{(1)}\beta_{2}}{4\beta_{1}^{3}}\hat{q}\left[
\frac{\ln(2\hat{q})+1}{\hat{q}}-\frac{\ln(2\hat{b})+1}{\hat{b}}\right]
+\frac{A^{(1)}\beta_{2}}{8\beta_{1}^{3}}\left[
\ln^{2}(2\hat{q})-\ln^{2}(2\hat{b})\right]
\nonumber \\
&&+\frac{A^{(1)}\beta_{2}}{8\beta_{1}^{3}}
\ln\left(\frac{e^{2\gamma_E-1}}{2}\right)\left[
\frac{\ln(2\hat{q})+1}{\hat{q}}-\frac{\ln(2\hat{b})+1}{\hat{b}}\right]
-\frac{A^{(2)}\beta_{2}}{16\beta_{1}^{4}}\left[
\frac{2\ln(2\hat{q})+3}{\hat{q}}-\frac{2\ln(2\hat{b})+3}{\hat{b}}\right]
\nonumber \\
& &-\frac{A^{(2)}\beta_{2}}{16\beta_{1}^{4}}
\frac{\hat{q}-\hat{b}}{\hat{b}^2}\left[2\ln(2\hat{b})+1\right]
+\frac{A^{(2)}\beta_{2}^2}{432\beta_{1}^{6}}
\frac{\hat{q}-\hat{b}}{\hat{b}^3}
\left[9\ln^2(2\hat{b})+6\ln(2\hat{b})+2\right]
\nonumber \\
&& +\frac{A^{(2)}\beta_{2}^2}{1728\beta_{1}^{6}}\left[
\frac{18\ln^2(2\hat{q})+30\ln(2\hat{q})+19}{\hat{q}^2}
-\frac{18\ln^2(2\hat{b})+30\ln(2\hat{b})+19}{\hat{b}^2}\right]
\label{sss}
\end{eqnarray}
\end{widetext}
where $\hat q$ and $\hat b$ are defined by
\begin{equation}
{\hat q} \equiv  {\rm ln}\left(xQ/(\sqrt 2\Lambda_{QCD})\right),~
{\hat b} \equiv  {\rm ln}(1/b\Lambda_{QCD})
\end{equation}
The coefficients $\beta_{i}$ and $A^{(i)}$ are
\begin{eqnarray}
& &\beta_{1}=\frac{33-2n_{f}}{12}\;,\;\;\;\beta_{2}=\frac{153-19n_{f}}{24}\; ,
A^{(1)}=\frac{4}{3}\;,
\nonumber \\
& & A^{(2)}=\frac{67}{9}-\frac{\pi^{2}}{3}-\frac{10}{27}n_
{f}+\frac{8}{3}\beta_{1}\ln\left(\frac{e^{\gamma_E}}{2}\right)\;
\end{eqnarray}
and $\gamma_E$ is Euler constant.


\end{document}